\documentclass[aps,prd,10pt,twocolumn,superscriptaddress,nofootinbib,showkeys,showpacs,altaffilletter]{revtex4-1}

\usepackage{graphicx}
\usepackage{dcolumn}
\usepackage{amssymb}
\usepackage{amsmath}
\usepackage{amsfonts}
\usepackage{amsbsy}
\usepackage{color}
\usepackage{rotating}
\usepackage[english]{babel}
\usepackage{multirow}
\usepackage{hyperref}

\def\be {\begin{equation}}
\def\ee {\end{equation}}
\def\bea {\begin{eqnarray}}
\def\eea {\end{eqnarray}}
\def\ba {\begin{align}}
\def\ea {\end{align}}

\newcommand{\dd}{{\rm d}}
\newcommand{\weff}{w_{\rm eff}}

\newcommand{\lsim}   {\mathrel{\mathop{\kern 0pt \rlap
  {\raise.2ex\hbox{$<$}}}
  \lower.9ex\hbox{\kern-.190em $\sim$}}}
\newcommand{\gsim}   {\mathrel{\mathop{\kern 0pt \rlap
  {\raise.2ex\hbox{$>$}}}
  \lower.9ex\hbox{\kern-.190em $\sim$}}}

\begin{document}

\title{Observational constraints on cosmological future singularities}

\author{Jose Beltr\'an Jim\'enez}
\affiliation{Aix Marseille Univ, Universit\'e de Toulon, CNRS, CPT, Marseille, France}
\author{Ruth Lazkoz}
\affiliation{Fisika Teorikoaren eta Zientziaren Historia Saila, Zientzia eta Teknologia Fakultatea,\\
Euskal Herriko Unibertsitatea, 644 Posta Kutxatila, 48080 Bilbao, Spain}
\author{Diego S\'aez-G\'omez}
\affiliation{Instituto de Astrof\'isica e Ci\^encias do Espa\c{c}o, Departamento de F\'isica, Faculdade de Ci\^encias da Universidade de Lisboa, Edif\'icio C8, Campo Grande, P-1749-016
Lisbon, Portugal}
\author{Vincenzo Salzano}
\affiliation{Institute of Physics, University of Szczecin, Wielkopolska 15, 70-451 Szczecin, Poland}

\date{\today}

\begin{abstract}
In this work we consider a family of cosmological models featuring future singularities. This type of cosmological evolution is typical of dark energy models with an equation of state violating some of the standard energy conditions (e.g. the null energy condition). Such kind of behavior, widely studied in the literature, may arise in cosmologies with phantom fields, theories of modified gravity or models with interacting dark matter/dark energy. We briefly review the physical consequences of these cosmological evolution regarding geodesic completeness and the divergence of tidal forces in order to emphasize under which circumstances the singularities in some cosmological quantities correspond to actual singular spacetimes. We then introduce several phenomenological parameterizations of the Hubble expansion rate to model different singularities existing in the literature and use SN Ia, BAO and $H(z)$ data to constrain how far in the future the singularity needs to be (under some reasonable assumptions on the behaviour of the Hubble factor). We show that, for our family of parameterizations, the lower bound for the singularity time can not be smaller than about $1.2$ times the age of the universe, what roughly speaking means $\sim 2.8$ Gyrs from the present time.

\end{abstract}
\date{\today}

\maketitle

\section{Introduction}
The standard model of cosmology together with the inflationary paradigm provide an accurate description of the universe, although it requires the presence of three unknown ingredients, namely: Dark matter, dark energy and the inflaton field. The last two share the property of being introduced in order to support phases of accelerating expansion. Moreover, while the inflaton accounts for the first instants of life of our universe, dark energy should determine its final fate as the component that will eventually dominate. If dark energy turns out to be simply a cosmological constant, then we are doomed to an asymptotically de Sitter universe in the future. The situation is much more subtle when dynamical dark energy or modified gravity is brought in as possible explanations for the late time accelerated expansion (for a review about dark energy models, see \cite{Nojiri:2010wj}). In some cases, dark energy is ascribed to a so-called phantom fluid, i.e., a fluid satisfying $\rho+p<0$ and, thus, violating the Null Energy Condition (NEC) \cite{phantom}. For a set of minimally coupled scalar fields, this condition implies the presence of, at least, a Laplacian instability in the inhomogeneous perturbations, although this can be resolved by allowing non-minimal couplings (see for instance \cite{NECscalars}). Moreover, such kind of behavior can be also a consequence of a modification of General Relativity instead of a fluid with a non-standard equation of state \cite{FRgravity}.  In any case, the phantom behavior may affect the background evolution giving rise to a future singularity occurring at a finite time where the scale factor diverges.  Nevertheless, note that some models with violations of the null energy condition do not drive the universe to a singularity but to regular scenarios that may affect the local structures, known as \textit{little Rip}, \textit{Pseudo-Rip} and \textit{Little Sibling} \cite{Frampton:2011sp,Frampton:2011aa,Bouhmadi-Lopez:2014cca}.

The described singular behaviour is actually shared by many dynamical dark energy models and modified gravity scenarios, where divergences in different cosmological parameters at a finite time can appear. The nature of the future singularities may differ among the different scenarios and they can be classified according to the cosmological parameters that diverge. An alternative way of classifying the future singularities is by means of the derivative of the scale factor that diverges. This classification is very useful because it helps understanding the severity of the different types of singularities (for a classification of cosmological singularities, see Ref.~\cite{Nojiri:2005sx,Fernandez-Jambrina:2014sga}). At this respect, it is worth reminding that a singular spacetime is characterized by the incompleteness of the geodesics \cite{HawkingEllis}. Since the geodesic equations are linear in the connection, it will contain, at most, first derivatives of the metric. Thus, the geodesics will be regular as long as the metric is continuous at the singularity. For a cosmological model, this will mean that the scale factor should remain finite at the singularity, even if divergences in the Hubble expansion rate or its derivatives are present. This type of behaviour has been recently used in \cite{Tilquin:2015gza} in order to replace the Big Bang singularity with a milder one that can be trespassed by the geodesics.

Another useful equation in order to characterize the strength of a singularity is the geodesics deviation equation. That equation essentially determines the tidal forces suffered by two infinitesimally close geodesics and it depends on the curvature of the spacetime. This means that tidal forces are sensitive to singularities which do not necessarily affect the completeness of the geodesics. Again in a cosmological context, if the scale factor remains regular, but the Hubble rate diverges, it is possible to have a regular geodesic congruence with divergent tidal forces. Some criteria based on the behaviour of the Riemann tensor as we approach the singularity exist in the literature to decide whether the singularity is strong or weak, being the Tipler \cite{Tipler:1977zza} and Krolak \cite{Krolak} conditions two widely used ones.

Regardless the physical consequences of having a future singularity at a finite time, a natural question to ask is how close a given type of singularity is to us \cite{Caldwell:2003vq}. This is the analogous of asking about the age of the universe, determined by our distance to the original Big Bang singularity. Nevertheless, in the same way as we do not expect the Big Bang singularity to exist actually, but rather being regularized by some quantum effects, high curvature corrections to  Einstein's gravity or even by varying physical constants \citep{DabrowskiVar}, we do not expect the future singularities to be physical, at least the strongest types where physical quantities diverge \cite{BouhmadiLopez:2004me}. However, it will be useful to have some estimation on how close to us a given singularity can be and, therefore, have an idea of how far in the future we could extrapolate a model with a certain type of future singularity. It is important to notice that an effective equation of state for dark energy $w<-1$ is within the confidence regions of observational data \cite{Lazkoz:2006gp} so the possibility of having a future singularity is plausible. Moreover, such models have also received attention because of some theoretical implications, since possible quantum effects close to the singularity become important. We know that General Relativity is to be regarded as an effective field theory whose strong coupling scale is, in the most optimistic scenario, at the Planck scale. Thus, knowing at which time the singularity is essentially reached will give us also an idea of until when we can keep using General Relativity as an effective field theory.

The purpose of the present work is precisely to draw such an estimation in a fairly model independent framework. An important difficulty arising here with respect to the Big Bang case is that, while in that case we have control on the different phases that the universe has gone trough from the initial singularity until today, for the future singularity we cannot know what the future phases will be. Thus, we need to make some assumptions to eventually determine how close the singularity can be. In order to achieve this, we will use some classes of phenomenological parameterizations for the Hubble expansion rate as proxies for a universe with a transition from  a matter dominated era to a dark energy phase leading to a future singularity. We will then confront them to SN Ia, BAO and $H(z)$ data to obtain the time of the singularity. Obviously, there could be transient phases that could delay the singularity, but this will not concern us since we are actually interested in obtaining a general lower bound for a future singularity.

The paper is organized as follows: section \ref{Sect2} is devoted to a brief review about future cosmological singularities. In section \ref{Sect3}, the parameterizations of the Hubble rate which are analyzed in the paper are introduced. Then, the observational data used to fit the models is described in section \ref{Sect4}. Finally, section \ref{Sect5} is devoted to the results discussions.

\section{Future cosmological Singularities}
\label{Sect2}

Assuming a homogeneous and isotropic universe at large scales, in compliance with the cosmological principle, the metric is given by the Friedmann-Lema\^itre-Robertson-Walker (FLRW) line element which is expressed as follows
\be
ds^2=-dt^2+a(t)^2\left(dx^2+dy^2+dz^2\right)\ ,
\label{FLRWmetric}
\ee
where we have assumed spatial flatness. Within General Relativity and assuming a perfect fluid as matter source, the gravitational equations can be written as
\be
H^2=\frac{8\pi G}{3}\rho\ ,\quad \dot{H}=-4\pi G (\rho+p)\ .
\label{FLRWeq}
\ee
Here $\rho$ and $p$ are the energy and pressure densities respectively of the perfect fluid, while $H=\frac{\dot{a}}{a}$ is the Hubble parameter. These equations are enough to describe the background cosmological evolution once the matter content of the universe is specified. In addition to these equations, the Bianchi identities allow to obtain the continuity equation $\dot{\rho}+3H(1+w)\rho=0$ with $w\equiv p/\rho\neq-1$ the equation of state (EoS) parameter. In cosmological scenarios based on non-standard fields, general fluids of modified gravity, several singularities have been found to appear at a future finite cosmic time. The different types of finite late-time singularities can be classified according to the divergent cosmological quantity at the singularity as follows (see Refs.~\cite{Nojiri:2005sx,Fernandez-Jambrina:2014sga},
\begin{itemize}
\item Type I (``Big Rip singularity''): For $t\rightarrow t_s$, $a\rightarrow \infty$ and $\rho\rightarrow \infty$, $|p|\rightarrow \infty$. Time-like geodesics are incomplete \cite{Caldwell:2003vq,FernandezJambrina:2006hj}.
\item Type II (``Typical Sudden singularity''): For $t\rightarrow t_s$, $a\rightarrow a_s$ and $\rho\rightarrow \rho_s$, $|p|\rightarrow \infty$. Geodesics are not incomplete. This is classified as a weak singularity (see Ref.~\cite{barrow1}).
\item Type III (``Big freeze''): For $t\rightarrow t_s$, $a\rightarrow a_s$
and $\rho\rightarrow \infty$, $|p|\rightarrow \infty$. No geodesics incompleteness. They can be weak or strong (see Ref.~\cite{BouhmadiLopez:2006fu}).
\item Type IV (``Generalized Sudden singularity''): For $t\rightarrow t_s$, $a\rightarrow a_s$ and
$\rho\rightarrow \rho_s$, $p \rightarrow p_s$ but higher derivatives of Hubble parameter diverge. They are weak singularities \cite{Dabrowski:2013sea}.
\item Type V (``$w$-singularities''):  For $t\rightarrow t_s$, $a\rightarrow \infty$ and $\rho\rightarrow 0$, $|p|\rightarrow 0$ and $w=p/\rho\rightarrow \infty$. These singularities are weak (see Ref.~\cite{Dabrowski:2009kg}).
\end{itemize}
Type III singularities were shown to appear naturally in traditional vector-tensor theories of gravity \cite{Jimenez:2008au}, while type II singularities can appear in a novel class of vector field theories that arise in generalized Weyl geometries \cite{Jimenez:2016opp}. In addition, there are other scenarios where no quantity diverges at a finite time but at infinity, namely the ``Little Rip'' \cite{Frampton:2011sp}, ``Pseudo-Rip'' \cite{Frampton:2011aa} and ``Little Sibling'' \cite{Bouhmadi-Lopez:2014cca}. The above classification is useful since it groups together different models exhibiting a background evolution where some cosmological quantity meets a divergence in the future. The fact that some given quantities might have a divergence is usually regarded as a non-desirable feature to have in a regular spacetime. However, a regular spacetime is only defined in terms of its geodesic completeness. Thus, a spacetime with a curvature divergence can be regular as long as the geodesics can {\it smoothly} go through the divergence. Hence, cosmological models with some of the divergences in the above classification do not need to correspond to singular spacetimes and, consequently, singular future universes. In order to study whether the different singularities correspond to a geodesically incomplete spacetime we will consider the geodesic equations given by
\be
\frac{\dd x^\mu}{\dd \lambda^2}+\Gamma^\mu_{\alpha\beta}\frac{\dd x^\alpha}{\dd \lambda}\frac{\dd x^\beta}{\dd \lambda}=0
\ee
where $\lambda$ is some affine parameter (proper time for instance for non-null geodesics) and $\Gamma^\mu_{\alpha\beta}$ are the corresponding Christoffel symbols. This equation already shows that it is the connection which determines the smoothness of the geodesics. In general, the solutions of the differential equations will be better behaved than the coefficients of the equations, so it is plausible to have a divergence in the connection with the geodesics remaining well-defined. It is also important to notice that the curvature contains derivatives of the connection and, therefore, there can be situations with curvature divergences, but where the connection (and consequently the geodesics) are perfectly regular. We will illustrate this below for some specific cases. The relevant case for the cosmological evolution is a spacetime described by the FLRW metric. In that case, the geodesic equations read\footnote{Here we will focus on spatially-flat universes. For the general case see \cite{FernandezJambrina:2004yy}}
\bea
\frac{\dd^2 t}{\dd \lambda^2}+Ha^2\delta_{ij}\frac{\dd x^i}{\dd \lambda}\frac{\dd x^j}{\dd \lambda}=0\, ,\label{tgeodesic}\\
\frac{\dd^2 x^i}{\dd \lambda^2}+2H\frac{\dd x^i}{\dd \lambda}\frac{\dd t}{\dd \lambda}=0\label{xgeodesic}.
\eea
These equations can be easily integrated. We start by rewriting the Hubble parameter in terms of the affine parameter as
\be
H=\frac{\dot{a}}{a}=\frac{1}{a}\frac{\dd a/\dd \lambda}{\dd t/\dd\lambda}.
\ee
Then, we can rewrite Eq. (\ref{xgeodesic}) as
\be
\frac{\dd}{\dd \lambda}\left(a^2\frac{\dd x^i}{\dd \lambda}\right)=0
\ee
which can be immediately integrated to obtain
\be
\frac{\dd x^i}{\dd \lambda}=\frac{u^i_0}{a^2}
\ee
with $u^i_0$ some integration constants. We can then use this solution into Eq. (\ref{tgeodesic}) to obtain
\be
\left(\frac{\dd t}{\dd\lambda}\right)^2=\frac{\vert\vec{u}_0\vert^2}{a^2}+C_0
\ee
where $C_0$ is another integration constant. We thus see that the geodesics will be regular (with a well-defined tangent vector) as long as the scale factor remains regular. If the scale factor does not diverge and is non-vanishing (so the metric is regular) the 4-velocities of the geodesics remain regular and the spacetime will be said to be non-singular. If the scale factor diverges at some point, then the geodesics stop there and cannot go through it. As we have discussed above, it is important to notice that the geodesics are insensitive to divergences in the expansion rate $H$ or its derivatives if they do not correspond to a singular behavior of the scale factor. This will be the case of the types II, III and IV singularities in the above classification where the scale factor remains finite while all the divergences only appear in its derivatives.

So far we have discussed the singularities from the point of view of geodesic completeness. Another class of criteria that is useful to study the presence of a singular physical behaviour is the geodesic deviation equation, which allows to infer the potential existence of divergences in tidal forces. The corresponding equations depend on the Riemann tensor, which explicitly contains the Hubble expansion rate and its first time-derivative so that it is, in principle, sensitive to divergences that do not affect the geodesics themselves. Two common criteria to classify these divergences are the so-called strong curvature divergences in the Tipler and Krolak sense, which are respectively characterized by the following integrals:
\bea
T(u)&\equiv&\int_0^\lambda \dd\lambda' \int_0^{\lambda'} \dd \lambda''R_{ij}u^i u^j,\\
K(u)&\equiv&\int_0^\lambda \dd{\lambda'} ''R_{ij}u^i u^j.
\eea
with $u^i$ the 4-velocity of the geodesic approaching the singularity. Here again we see that divergences in the curvature do not necessarily lead to a physical singularity because integrals of a given function are generally better behaved than the function itself. Thus, even if the spacetime contains a curvature divergence, it can remain {\it regular} according to the above criteria. The physical reason roots in the fact that the geodesic deviation equation measures the infinitesimal deviation, i.e., the tidal force between infinitesimally closed geodesics. However, extended physical objects have a finite physical volume and the above criteria precisely give the conditions for a finite volume to remain finite when going through the singularity. On the other hand, if the tidal forces are strong enough such that the volume shrinks to zero, the singularity is said to be strong.

\section{The models}
\label{Sect3}

In this section we will describe the parameterizations that we will use for the subsequent confrontation to observational data. We emphasize that we intend to establish a general lower bound for the time of the future singularity $t_s$. Since we are dealing with future singularities occurring at a finite proper time, it is reasonable to perform our parameterizations in terms of proper time. Moreover, as we have discussed, the severity of the different types of singularities is essentially determined by whether the scale factor or any of its time-derivatives presents a divergence. Therefore, the natural cosmological quantity to parameterize is the scale factor. However, for convenience when confronting to SN Ia and BAO, it will be more appropriate to parameterize the Hubble expansion rate directly. By doing this, we also avoid the ambiguity in the normalization of the scale factor.

As commented in the introduction, the main difficulty with respect to constraining the time of the Big Bang is that, while we have an accurate knowledge about the past history of the universe so we can robustly compute such a time, the future evolution of the universe is completely unknown. Because of that we need to make some relatively strong assumptions on our parameterizations. First of all, we want to have an approximate matter dominated phase at early times; we will use low-redshift ($z\leq2$) cosmological data so that by early time we actually mean well inside the matter domination epoch, but much later than equality and decoupling times, i.e., redshifts $10\leq z\leq1000$. In order to comply with this requirements we propose to use the following form for the Hubble expansion rate:
\be
H(t)= \frac{2}{3 t}+F(t,t_s)\ ,
\label{HGen}
\ee
where $t_s$ is the time when one of the above singularities occur and the function $F(t,t_s)$ is assumed to be negligible for $t\ll t_0$ with $t_0$ the present time, such that $H(t\ll t_0)\simeq \frac{2}{3 t}$ as it corresponds for a matter dominated universe. In terms of the scale factor, this translates into a parameterization of the form
\be
a(t)\propto g(t,t_s)t^{2/3}\quad {\rm with}\quad F(t,t_s)=\frac{\dot{g}}{g}.
\ee
This matter dominated phase will then be matched to an evolution with a future time singularity, i.e., $F(t,t_s)$ is a function that either itself or some of its time-derivatives diverges at $t=t_s$. We will assume that this divergence originates from the fact that the differential equation of the underlying physical model presents a regular singular point so that the solution near the singularity can be expressed as a Frobenius series. We will further assume that the transition from the matter era is sufficiently fast so that the dominant term of the series rapidly takes over. This will not affect our goal of obtaining a lower bound for the time of the transition since making the transition slower typically delays the appearance of the divergence.
Since we are looking for a future divergence where a given derivative of the scale factor diverges while the lower derivatives remain finite, a reasonable Ansatz for $F(t,t_s)$ is some half-integer power. With these considerations in mind, we have chosen the specific parameterizations summarized in Table.~\ref{tab:sing_classification} together with their main properties\footnote{We have tested other parameterizations for each type of singularity and found similar results, so we only report here the results for these representative parameterizations.}.  All the models contain two parameters characterizing the time of the singularity $t_s$ and an additional parameter $n$ that regulates the time of the transition from matter domination. Notice that all the parameterizations share the property of containing a late-time de Sitter evolution when the time of the singularity is sent to the asymptotic future\footnote{For the model $C$ we need to simultaneously send $n$ to infinity so that the product $n\log(1-t/t_s)$ remains finite.} $t_s\rightarrow \infty$. However, it is important to notice that the existence of a matter phase at early times matching a de Sitter universe in the asymptotic future does not necessarily mean that the evolution mimics that of a $\Lambda$CDM model, because the transition era between the two phases may be completely different. In fact, it is not difficult to see that none of our parameterizations contains $\Lambda$CDM within its parameter space.


Although our parameterizations do not rely on a specific gravitational theory, in order to make contact with previous literature and give a physical intuition of what kind of theoretical models might be described by our parameterizations, we will now consider some explicit cases. If we assume General Relativity for the gravitational interaction, we can define an effective equation of state for the universe as
\be
\weff\equiv\frac{p}{\rho} = -1 -\frac{2}{3} \frac{\dot{H}}{H^2}.
\label{eq:weff}
\ee
Thus, we can interpret our parameterizations in terms of an effective equation of state parameter for the content of the universe, assuming a perfect fluid form and a barotropic equation of state. Since at early times our parameterizations give $H\simeq 2/(3t)$, we recover a matter dominated universe with $\weff\simeq 0$ as it should.

On the other hand, the models considered in this manuscript can alternatively be interpreted as the result of the interaction between dust and dark energy. In such a case, the continuity equations for both fluids would be coupled, and could be expressed by \cite{Nojiri:2005sx},
\begin{eqnarray}\label{Eqbis1}
\dot{\rho}_m+3H\rho_m&=&Q(t)\ , \nonumber \\
\dot{\rho}_{DE}+3H(1+w_{DE})\rho_{DE}&=&-Q(t)\ .
\end{eqnarray}
Here $Q(t)$ accounts for the energy exchange between both fluids. By combining equations (\ref{Eqbis1}) with the Friedmann equation, the following expression for $Q(t)$ is obtained:
\begin{eqnarray}\label{Eqbis2}
Q(t) &=& \frac{1}{\kappa^2 w_{DE}} \left[9(1+w_{DE})H^3+6(2+w_{DE})H\dot{H}+2\ddot{H} \right. \nonumber \\
&-& \left.  \frac{\dot{w}_{DE}}{w_{DE}} \left( 3 H^2 + 2 \dot{H}\right) \right]
\end{eqnarray}
Hence, by assuming a particular Ansatz $H=H(t)$, the corresponding interacting term $Q(t)$ is obtained. Moreover, note that other type of non-interacting models can lead to the concerned models of Table \ref{tab:sing_classification} by means of non-standard EoS's for dark energy that effectively stand for modifications of the Hilbert-Einstein action, viscosity terms (see Ref.~\cite{Odintsov_DF}). Nevertheless, the later case may lead to negative energy densities and other undesirable consequences, as shown below for our parameterizations of the Hubble parameter.


Finally, we can also mention that a given background evolution for the scale factor can be mapped onto a scalar field theory by suitable choice of the action. In the spirit of the effective field theory of dark energy, we can think of the time coordinate as corresponding to a foliation of the spacetime according to the scalar field, where the unitary gauge has been chosen. Then, a natural interpretation of the future singularity would be a point where the scalar field meets a pathology in its evolution as dictated by the field equations.

{\renewcommand{\tabcolsep}{2.mm}
{\renewcommand{\arraystretch}{2}
\begin{table*}[ht!]
\begin{minipage}{\textwidth}
\centering
\resizebox*{\textwidth}{!}{
\begin{tabular}{|c|c|cc|ccccccc|}
\hline
$Label$ & $Type$ & $H(x)$ & $a(x)$  & $a$       & $H$         & $\dot{H}$ & $\ddot{H}$  & $\rho$     & $p$       & $\weff$  \\
\hline \hline
$A$ & $I$  & $\frac{2}{3 x} + \frac{2 n }{3\left(1-x/x_s\right)}$ &  $a_0x^{\frac{2}{3}}\left(x_{s}-x\right)^{-\frac{2}{3} \cdot n x_{s}}$    &  $\infty$ & $\infty$    & $\infty$  & $\infty$      & $\infty$   & $-\infty$ & $w_{s} < 0$ \\
\hline
$B$ & $III$ & $\frac{2}{3 x} + \frac{2 n}{3\sqrt{1-x/x_{s}}}$  &  $a_0x^{\frac{2}{3}} \exp{[-\frac{4}{3}  n \sqrt{x_{x}\left(x_{s}-x\right)}]}$      &  $a_{s}$  & $\infty$    & $\infty$  & $\infty$      & $\infty$   & $-\infty$ & $-\infty$  \\
\hline
$C$ & $III$ & $\frac{2}{3 x} - \frac{2 n}{3} \log \left(1-\frac{x}{x_{s}}\right)$ & $a_0x^{\frac{2}{3}} \exp{[-\frac{2}{3} n \left( x - x_{s}\right) \left(-1+\log \left[ 1- x/x_{s} \right] \right)]}$     &  $a_{s}$  & $\infty$    & $\infty$  & $\infty$    & $\infty$   & $-\infty$ & $-\infty$  \\
\hline
$D$ & $II$ & $\frac{2}{3 x} + \frac{2 n}{3} \sqrt{1-\frac{x}{x_{s}}}$   & $a_0\left(x/x_{s}\right)^{\frac{2}{3}} \exp{[-\frac{4}{9} \cdot n x_{s} \left( 1- x/x_{s}\right)^{\frac{3}{2}}]}$   &  $a_{s}$  & $H_{s} > 0$ & $-\infty$    & $-\infty$    & $\rho_{s}$ & $\infty$  & $\infty$   \\
\hline
$E$ & $IV$ & $\frac{2}{3 x} + \frac{2 n}{3} \left( 1 - \frac{x}{x_{s}}\right)^{3/2}$ & $a_0\left(x/x_{s}\right)^{\frac{2}{3}} \exp{[-\frac{4}{15} \cdot n x_{s} \left(1-x/x_{s} \right)^{\frac{5}{2}}]}$  &  $a_{s}$  & $H_{s} > 0$ & $\dot{H}_{s} < 0$    & $\infty$    & $\rho_{s}$ & $0$  & $0$ \\
\hline \hline
\end{tabular}}
\caption{In this table we summarize the 5 parameterizations that we propose to describe the different types of future singularities that we consider throughout this work. In the first column we give the label we will use for each case, while the second column indicates the type of singularity according to the classification in \cite{Nojiri:2005sx}. In the columns 3 and 4 we give the analytical expressions for $H(t)$ and $a(t)$ (where, as explained in the main text, the normalization $a_0$ must be chosen so that $a(x=1)=1$). In the last columns we give the behaviour of $a$, $H$ and some of its derivatives at the singularity. We also give the values of $\rho$, $p$ and $\weff$ for the theoretical interpretation discussed in Section III. It is important to keep in mind that those values depend on the underlying theoretical model and we only give them here for illustrative purposes.}\label{tab:sing_classification}
\end{minipage}
\end{table*}}}

\section{Data}
\label{Sect4}

The analysis has been performed using three different standard cosmological tools. They are at low redshift $(z \lesssim 2)$, because we are not interested in changing early time evolution and we assume that a possible signature for ``future'' evolution toward a singularity, if any, is detectable now or, at least, in the recent past only.

Just for sake of clarity and computational motivations, all the models we propose are written in terms of the dimensionless variable $x = t / t_{0}$, where $t_{0}$ is the age of the Universe. This means that all quantities will be measured in units of $t_0$ so, for instance, we will have
\begin{equation}
H(t) = \frac{H(x)}{t_{0}}.
\end{equation}
Therefore, in this case $t_{0}$ plays, in terms of fitting parameters, the role usually ascribed to the Hubble constant $H_{0}$ in the standard approach.

Since our parameterizations are explicitly expressed in terms of time, it will be more convenient to use all the standard integrals involved in the calculation of cosmological distances directly expressed as integrations over time, instead of transforming them into integrations over redshift, being the two of them related as
\begin{equation}
\int^{z}_{0} \frac{\dd\,\tilde{z}}{H(\tilde{z})} \quad \rightarrow \quad \int^{x}_{1} \frac{\dd\,\tilde{x}}{a(\tilde{x})}\; .
\end{equation}
The integrations over redshift are more convenient in the usual case because the observational data are given in terms of redshift. Thus, we will need to find the values of $x$ that correspond to the given redshifts, i.e., we need to find the functions $z=z(x)$ or, equivalently $a=a(x)$. This can be
easily obtained from the corresponding expression for $H(x)$ by solving the differential equation
\begin{equation}
H(x) = \frac{a'(x)}{a(x)} \;,
\end{equation}
where the prime stands for derivative with respect to $x$. This equation will be solved with the boundary condition $a(x=1)=1$, i.e., we normalize the scale factor to be 1 today. Thus we operationally define the time $t_{0}$ in our models by such condition. Notice that for all our parameterizations this can be done analytically. Therefore, we can obtain the values of $x_i$ corresponding to the measured values $z_i$ by numerically solving the equation $z_i=1/a(x_i)-1$.

\subsection{Hubble data from early-type galaxies}

We use the compilation of Hubble parameter measurements estimated with the differential evolution of passively evolving early-type galaxies as cosmic chronometers, in the redshift range $0<z<1.97$ and recently updated in \citep{Moresco}. The corresponding $\chi^2_{H}$ estimator is defined as
\begin{equation}\label{eq:hubble_data}
\chi^2_{H}= \sum_{i=1}^{24} \frac{\left( H(x_{i},\boldsymbol{\theta})-H_{obs}(x_{i}) \right)^{2}}{\sigma^2_{H}(x_{i})} \; ,
\end{equation}
with $\sigma_{H}(z_{i})$ the observational errors on the measured $H_{obs}(z_{i})$ values, and $\boldsymbol{\theta}$ is the vector of cosmological parameters, i.e., $(t_0, n, t_s)$ in our case. Moreover, we will add a gaussian prior, derived from the Hubble constant value given in \citep{Bennett}, $H_{0} = 69.6 \pm 0.7$. Notice that now $H_0$ is a derived quantity depending on the actual fitting parameters so
\begin{equation}
H_{0} = H_0(\boldsymbol{\theta}) = \frac{H(x=1,\boldsymbol{\theta})}{t_{0}}\; ,
\end{equation}
where the numerator $H(x=1,\boldsymbol{\theta})$ now depends on the parameters $n$ and $x_{s}$.

\subsection{Type Ia Supernovae}

We use the SN Ia data from the Union2.1 compilation \cite{Union21}. The $\chi^2_{SN}$ in this case is generally defined as
\begin{equation}
\chi^2_{SN} = \Delta \boldsymbol{\mathcal{F}}^{SN} \; \cdot \; \mathbf{C}^{-1} \; \cdot \; \Delta  \boldsymbol{\mathcal{F}}^{SN} \; ,
\end{equation}
with $\Delta\boldsymbol{\mathcal{F}}^{SN} = \mu_{theo} - \mu_{obs}$ the difference between the observed and theoretical value of the distance modulus $\mu$, the observable quantity for Union2.1 SN Ia, defined as:
\begin{equation}\label{eq:mu_union21}
\mu = 5 \log_{10} [d_{L}(z, \boldsymbol{\theta}) ] + \mu_{0} \; ;
\end{equation}
with $d_{L}(z)$ the dimensionless luminosity distance given by
\begin{equation}\label{eq:dl_H}
d_{L}(z, \boldsymbol{\theta}) = (1+z) \ \int_{0}^{z} \frac{\mathrm{d}\tilde{z}}{E(\tilde{z}, \boldsymbol{\theta})} \, ,
\end{equation}
where $E(z) = H(z)/H_{0}$ is the dimensionless Hubble function; $\mu_{0}$ a nuisance parameter combining the Hubble constant $H_{0}$ (or $t_{0}$ in our case) and the absolute magnitude of a fiducial SN Ia. As usual, we marginalize the $\chi^2_{SN}$ over $\mu_{0}$. Finally, $\mathbf{C}$ is the covariance matrix. In terms of integration over time, the dimensionless luminosity distance can be expressed as:
\begin{equation}
d_{L}(x, \boldsymbol{\theta}) = \frac{1}{a(x,\boldsymbol{\theta})} \int_{1}^{x} \frac{\mathrm{d}\tilde{x}}{a(\tilde{x}, \boldsymbol{\theta})} \, .
\end{equation}

{\renewcommand{\tabcolsep}{2.mm}
{\renewcommand{\arraystretch}{1.5}
\begin{table*}[ht!]
\begin{minipage}{0.95\textwidth}
\centering
\resizebox*{\textwidth}{!}{
\begin{tabular}{|c|ccc|cc|c|cc|}
\hline
$id.$ & \multicolumn{3}{c|}{$\Omega_m$} & $H_{0}$ & $t_{0}$ & $w_{\mathrm{eff},0}$ & $\mathcal{B}_{ij}$ & $\log \mathcal{B}_{ij}$\\
      & \multicolumn{3}{c|}{} & km s$^{-1}$ Mpc$^{-1}$ & Gyr & & & \\
\hline
$\Lambda CDM$ & \multicolumn{3}{c|}{$0.30^{+0.02}_{-0.02}$} & $69.6^{+0.7}_{-0.7}$ & $13.54^{+0.31}_{-0.29}$ & $-0.70$ & $1$ & $0$ \\
\hline \hline
$id.$ & $\Omega_m$ & $w_{0}$ & $w_{a}$ & $H_{0}$ & $t_{0}$ & $w_{\mathrm{eff},0}$ & $\mathcal{B}_{ij}$ & $\log \mathcal{B}_{ij}$\\
      &  & & & km s$^{-1}$ Mpc$^{-1}$ & Gyr & & & \\
\hline
$CPL$ & $0.36^{+0.05}_{-0.09}$ & $-1.00^{+0.23}_{-0.24}$ & $-1.14^{+1.87}_{-3.08}$ & $69.5^{+0.7}_{-0.7}$ & $13.25^{+0.38}_{-0.32}$ & $-0.64$ & $0.56$ & $-0.63$ \\
\hline \hline
$id.$ & $n$ & $\alpha_{s}$ & $t_{s}/t_{0}$ & $1/t_{0}$ & $t_{0}$ & $w_{\mathrm{eff},0}$ & $\mathcal{B}_{ij}$ & $\log \mathcal{B}_{ij}$\\
      &     &  &  & km s$^{-1}$ Mpc$^{-1}$ & Gyr & &  & \\
\hline
\multicolumn{9}{|c|}{Uniform prior} \\
\hline
$A$ & $0.27^{+0.07}_{-0.06}$ & $0.31^{+0.16}_{-0.18}$ & $2.17^{+0.85}_{-0.42}$ & $69.9^{+1.1}_{-1.1}$ & $14.00^{+0.23}_{-0.21}$ & $-0.74$ & $0.62$ & $-0.48$\\
\hline
$B$ & $0.33^{+0.06}_{-0.06}$ & $0.46^{+0.19}_{-0.24}$ & $1.78^{+0.74}_{-0.34}$ & $69.6^{+1.1}_{-1.2}$ & $14.06^{+0.24}_{-0.22}$ & $-0.70$ & $0.57$ & $-0.56$\\
\hline
$C$ & $0.99^{+0.55}_{-0.38}$ & $<0.29$ & $>2.22$ & $71.1^{+1.1}_{-1.0}$ & $13.76^{+0.20}_{-0.21}$ & $-0.92$ & $0.34$ & $-1.07$\\
\hline
$D$ & $0.66^{+0.08}_{-0.06}$ & $<0.24$ & $>2.43$ & $68.0^{+1.1}_{-1.2}$ & $14.39^{+0.25}_{-0.24}$ & $-0.48$ & $0.04$ & $-3.23$ \\
\hline
$E$  & $0.81^{+0.08}_{-0.07}$ & $<0.05$ & $>3.92$ & $67.4^{+1.1}_{-1.1}$ & $14.52^{+0.24}_{-0.23}$ & $-0.45$ & $0.01$ & $-4.37$\\
\hline
\multicolumn{9}{|c|}{Logarithmic prior} \\
\hline
$A$ & $0.30^{+0.08}_{-0.07}$ & $<0.32$ & $>2.15$ & $69.8^{+2.0}_{-2.2}$ & $14.02^{+0.46}_{-0.39}$ & $-0.78$ & $1.01$ & $0.007$\\
\hline
$B$ & $0.34^{+0.07}_{-0.07}$ & $0.42^{+0.21}_{-0.24}$ & $1.86^{+0.84}_{-0.40}$ & $69.4^{+1.7}_{-1.4}$ & $14.10^{+0.37}_{-0.35}$ & $-0.69$ & $1.21$ & $0.19$\\
\hline
$C$ & $1.44^{+0.67}_{-0.47}$ & $<0.12$ & $>3.16$ & $71.2^{+0.7}_{-1.0}$ & $13.75^{+0.20}_{-0.14}$ & $-0.86$ & $0.49$ & $-0.71$\\
\hline
$D$ & $0.60^{+0.04}_{-0.03}$ & $<0.05$ & $>3.92$ & $68.2^{+1.0}_{-1.0}$ & $14.34^{+0.21}_{-0.21}$ & $-0.53$ & $0.08$ & $-2.47$ \\
\hline
$E$ & $0.65^{+0.06}_{-0.04}$ & $<0.002$ & $>7.09$ & $68.0^{+1.0}_{-1.0}$ & $14.38^{+0.20}_{-0.21}$ & $-0.51$ & $0.06$ & $-2.82$\\
\hline \hline
\end{tabular}}
\caption{In this table we present the obtained results for the best fit of each parameterization.
In column 1 we give the label identifying each parameterization in Table \ref{tab:sing_classification}. In columns 2-5 we give the $1\sigma$ confidence levels for our primary model parameters (notice that they are different for the different cases and, in particular, $\Omega_m$ is not within the fitting parameters of our parameterizations). In column 6 we show the age of the Universe. We also show the effective equation of state parameter (as defined in \eqref{eq:weff}) for each parameterization evaluated at the present. Finally, in columns 8 and 9 we give the Bayesian evidence and ratio with respect to $\Lambda$CDM for Jeffreys' interpretation.}\label{tab:results}
\end{minipage}
\end{table*}}}

\subsection{Baryon Acoustic Oscillations}

We have also made use of baryon acoustic oscillations (BAO), in particular, the data collected in \citep{WiggleZ,SDSS12}\footnote{Data for SDSS DR12 release are available for download at \url{https://sdss3.org/science/boss_publications.php}.}. In this case the $\chi_{BAO}^2$ is defined as
\begin{equation}
\chi^2_{BAO} = \Delta \boldsymbol{\mathcal{F}}^{BAO} \; \cdot \; \mathbf{C}^{-1} \; \cdot \; \Delta  \boldsymbol{\mathcal{F}}^{BAO} \; ,
\end{equation}
where, as before, $\Delta\boldsymbol{\mathcal{F}}^{BAO} = F_{theo} - F_{obs}$ is the difference between the observed and theoretical value of the Alcock-Paczynski distortion parameter measured in a BAO survey, and defined as:
\begin{equation}\label{eq:bao_data}
F(z) = (1+z)D_{A}(z)\frac{H(z)}{c} \; ,
\end{equation}
with $c$ the speed of light, $H(z)$ the Hubble function, and $D_{A}$ the angular diameter given by:
\begin{equation}
D_{A}(z, \boldsymbol{\theta}) = \frac{c}{H_{0}(1+z)} \ \int_{0}^{z}
\frac{\mathrm{d}\tilde{z}}{E(\tilde{z}, \boldsymbol{\theta})} \, .
\end{equation}
Even in a standard scenario, the quantity $F(z)$ is independent of the parameter $H_{0}$ and can be written
\begin{equation}
F(z, \boldsymbol{\theta}) = \left( \int_{0}^{z}
\frac{\mathrm{d}\tilde{z}}{E(\tilde{z}, \boldsymbol{\theta})}\right) \cdot E(\tilde{z}, \boldsymbol{\theta})\;,
\end{equation}
which in our notation translates into
\begin{equation}
F(x, \boldsymbol{\theta}) = \left( \int_{1}^{x}
\frac{d\,x'}{a(x', \boldsymbol{\theta})}\right) \cdot \left(\frac{H(x', \boldsymbol{\theta})}{H(1, \boldsymbol{\theta})} \right)\;,
\end{equation}
which is independent of the parameter $t_{0}$. One important remark about BAO data concerns the possibility to use other than the Alcock-Paczynski variables like the angular diameter distance or other conveniently defined quantities (see, for example, \citep{WiggleZ,SDSS12}). Those quantities involve the calculation of the sound horizon at dragging epoch, which in turn requires knowledge about the density parameters of baryons and radiation. However, as explained in previous sections, our models only provide phenomenological parameterizations of the Hubble function in terms of cosmic time. For that we introduce some parameters whose relation to the more physical density parameters can only be inferred after assuming a particular theory of gravity.  Thus, as a very conservative choice, we have decided to perform our analysis using only the Alcock-Paczynski method. For sake of correctness, we have also to stress that even the Alcock-Paczynski quantities are derived making some assumptions related to the background cosmology (at least, at a very early stage of raw data analysis), as it is pointed out in \citep{WiggleZ2}. But, in the same reference, it is claimed that their final results are not very sensitive to the fiducial model.

Finally, the total $\chi^2$ to be minimized will be $\chi^2 = \chi^{2}_{H}+\chi^{2}_{H_{0}}+\chi^{2}_{SN}+\chi^{2}_{BAO}$. We minimize the total $\chi^2$ using the Markov Chain Monte Carlo (MCMC) method and we check its convergence with the method developed in \citep{Dunkley05}. In order to compare the models in the best statistical way possible, we have calculated the Bayesian evidence for each of them. The Bayesian evidence is defined as the probability of the data $D$ given the model $M$ with a set of parameters $\boldsymbol{\theta}$, $\mathcal{E}(M) = \int \mathrm{d}\boldsymbol{\theta} L(D|\boldsymbol{\theta},M)\pi(\boldsymbol{\theta}|M)$: $\pi(\boldsymbol{\theta}|M)$ is the prior on the set of parameters, normalized to unity, and $L(D|\boldsymbol{\theta},M)$ is the likelihood function.

We have been very careful in imposing priors; our parameters are, basically, $t_{0}$, $x_{s}$ and $n$. For numerical convenience, we have used the parameter $\alpha$ instead of $x_s$ defined as
\be
x_{s} = 1- \ln \alpha
\ee
in order to compactify the range $x_s\in(1,\infty)$ into $\alpha\in(0,1)$. Since we are compactifying the range of our parameter $x_s$ from an infinite range to a finite one, we face the question of imposing an appropriate prior.For the parameter $\alpha$, we have imposed two different priors on this range: a flat uniform prior, and a logarithmic one. This is due to the relation between $\alpha$ and the variable we are really interested in, $x_{s}$, so that we want to have a full sampling of the very low range for $\alpha$, which maps into $x_{s} \rightarrow \infty$ and, thus,  it corresponds to nearly non-singular scenarios (i.e., the singularity is pushed to the far infinity). For $n$ (and $t_{0}$) we assume a flat prior for only positive values, $n>0$, given that this is the condition to ensure present accelerated expansion for all the models. The only exception is for the singularity D, where acceleration is guaranteed for $n>\overline{n}(\alpha)>0$, with $\overline{n}(\alpha)$ numerically found imposing the condition $q(t)=0$, with $q(t)$ being the deceleration parameter. Thus, the parameters span sufficiently wide and general ranges in order to have the same weight for each model when calculating the Bayesian evidence. The evidence is estimated using the algorithm in \cite{evidence}; in order to reduce the statistical noise we run the algorithm many times obtaining a distribution of $\sim 100$ values from which we extract the best value of the evidence as the mean of such distribution.

In order to compare the goodness of the different parameterizations, we further calculate the Bayes Factor, defined as the ratio of evidences of two models, $M_{i}$ and $M_{j}$, $\mathcal{B}_{ij} = \mathcal{E}_{i}/\mathcal{E}_{j}$. If $\mathcal{B}_{ij} > 1$, model $M_i$ is preferred over $M_j$, given the data. We will use the $\Lambda$CDM model as the reference model $j$ (we have performed a further analysis with this model using the same data sets we have described above). The Bayesian evidence may be interpreted using Jeffreys' Scale \cite{jeffreys}, which tries to quantify the preference of a model against another based on the value of the evidence. In particular, if the $\ln \mathcal{B}_{ij}<1$, the evidence in favor of the highest-evidence model is not significant; if $1<\ln \mathcal{B}_{ij}<2.5$, the evidence is substantial; if $2.5<\ln \mathcal{B}_{ij}<5$, the evidence is strong; and if $\ln \mathcal{B}_{ij}>5$, the evidence is decisive. In \cite{nesseris}, it is shown that the Jeffreys' scale is not a fully-reliable tool for model comparison, but at the same time the statistical validity of the Bayes factor as an efficient model-comparison tool is not questioned: a Bayes factor $\mathcal{B}_{ij}>1$ unequivocally states that the model $i$ is more likely than model $j$. We present results in both contexts for reader's interpretation.

\section{Results}
\label{Sect5}

\begin{figure*}[htbp]
\centering
\includegraphics[width=6.3cm]{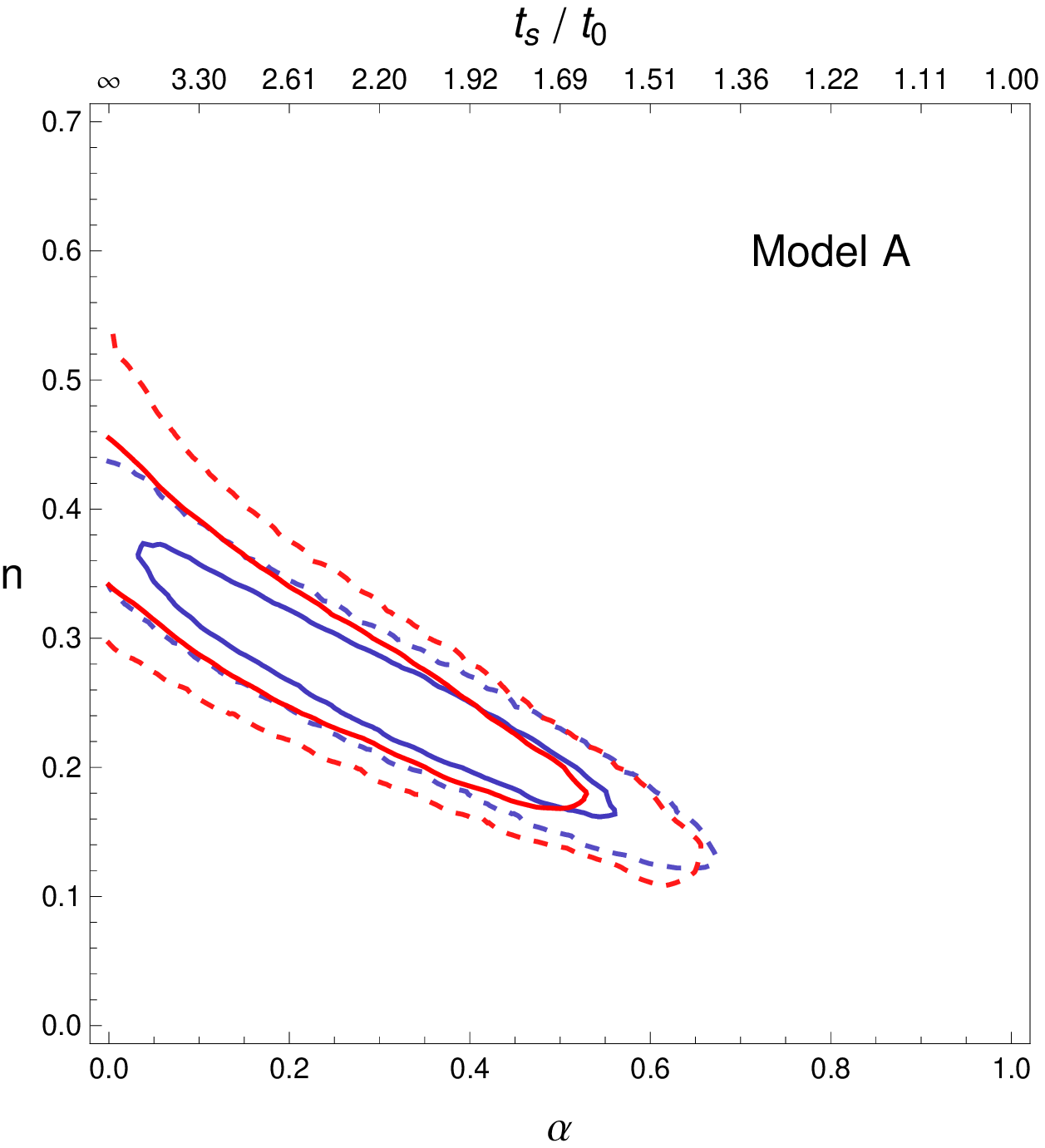}~~~
\includegraphics[width=6.3cm]{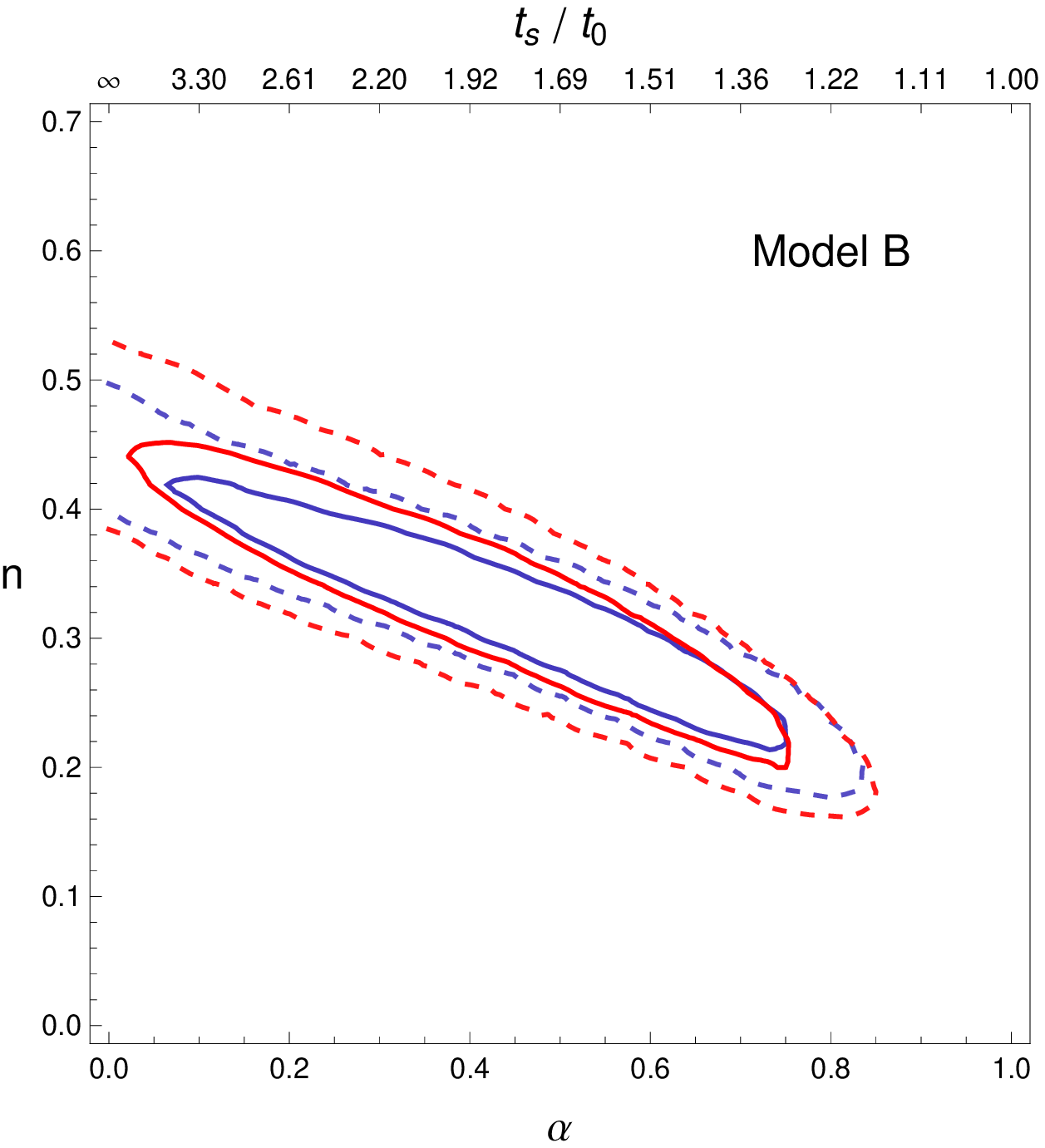}\\
~~~\\
\includegraphics[width=6.3cm]{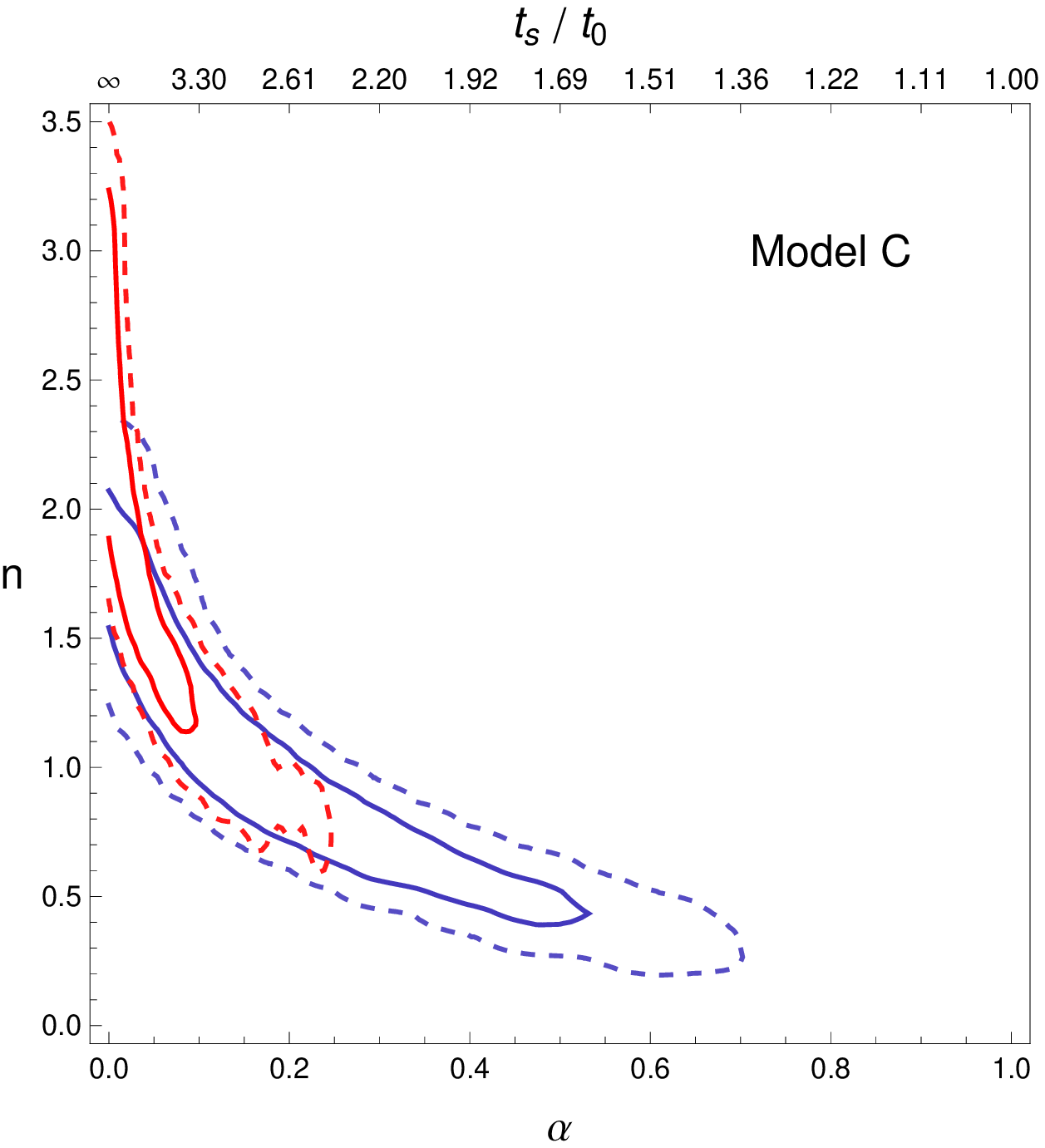}~~~
\includegraphics[width=6.3cm]{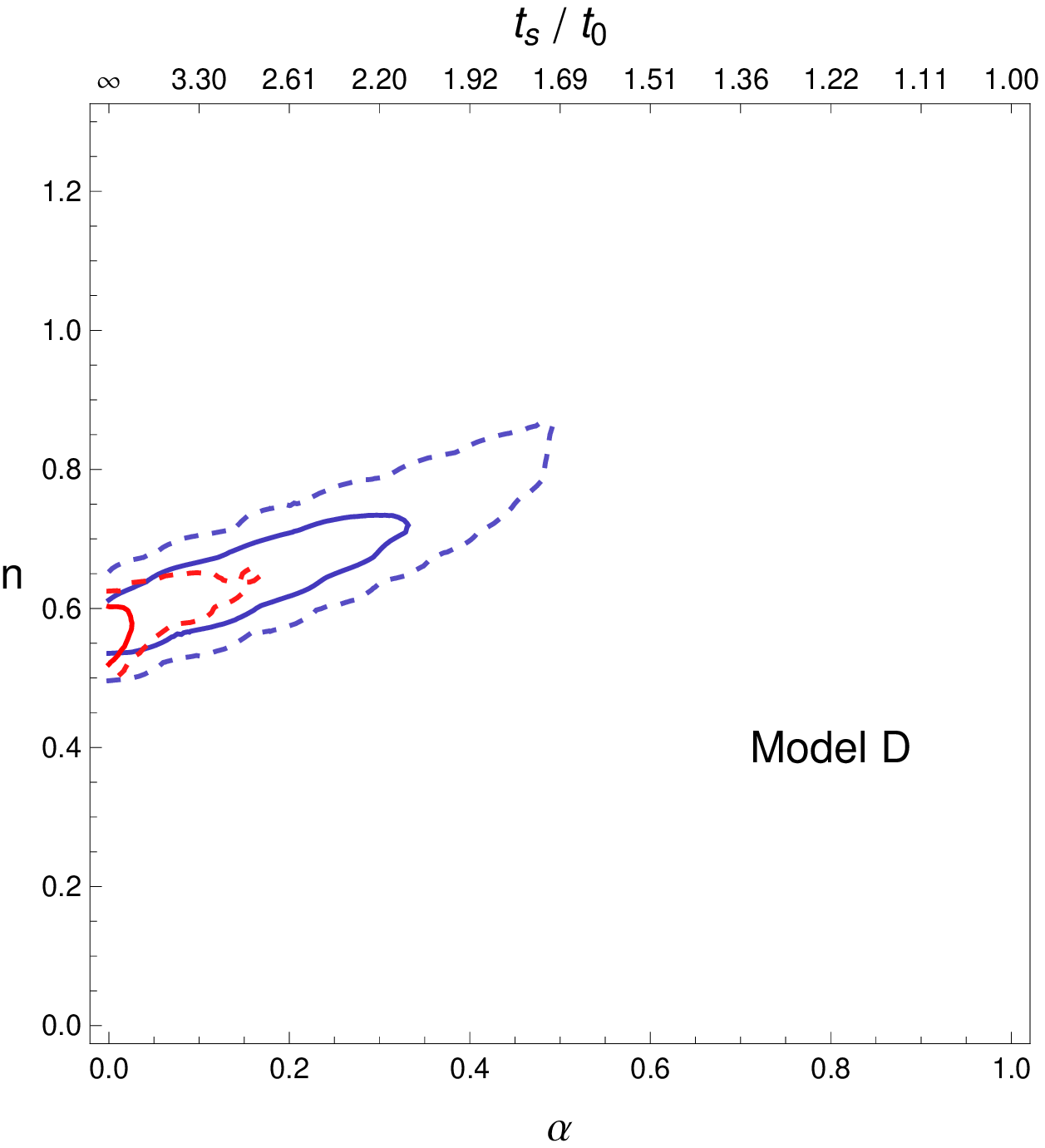}\\
~~~\\
\includegraphics[width=6.3cm]{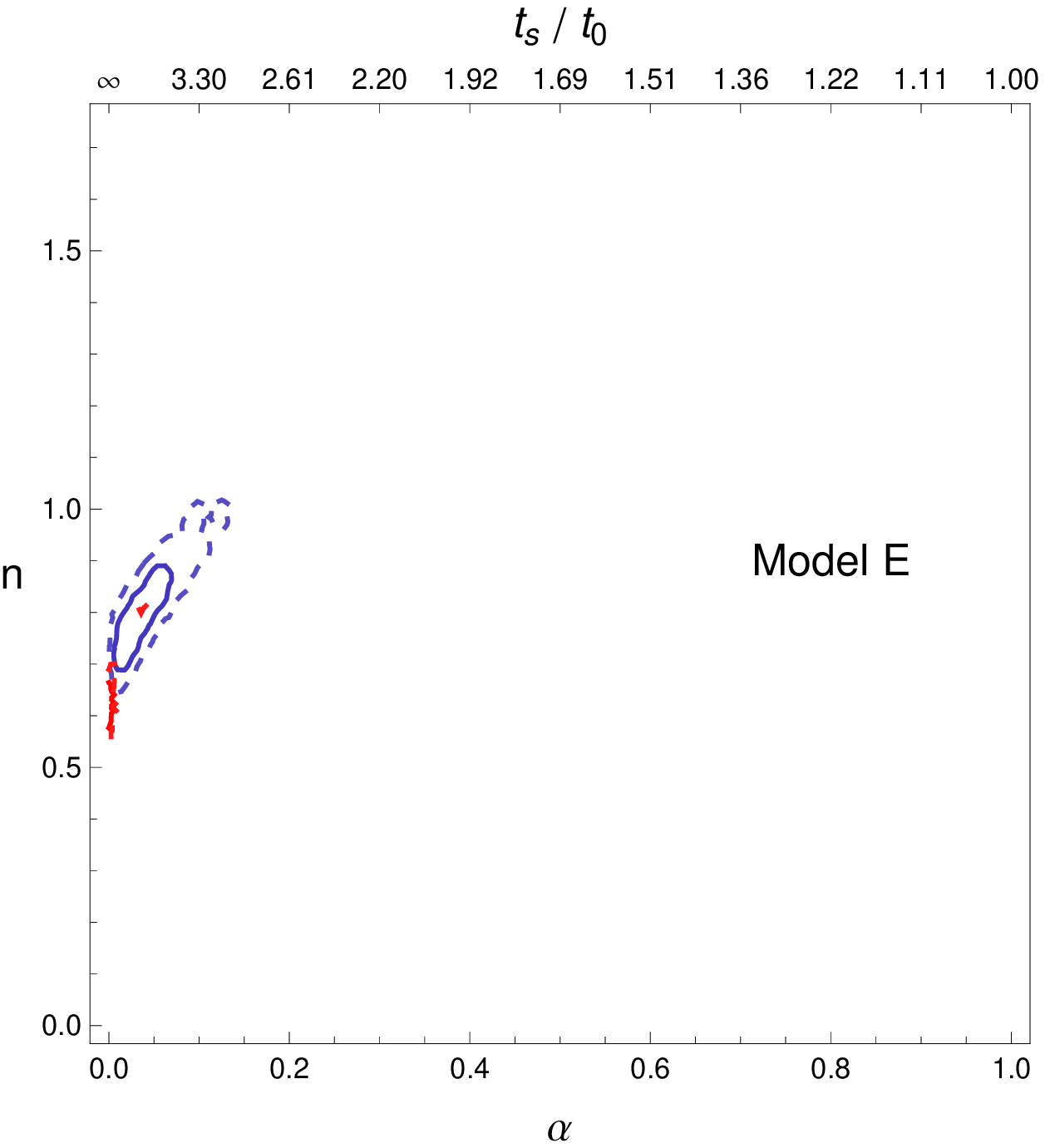}
\caption{Results: $68\%$ and $95\%$ confidences level for $n$ and $\alpha$. Blue: total combined data sets with uniform prior; red: total combined data sets with logarithmic prior. From left to right and from top to bottom: singularity A; singularity B; singularity C; singularity D; singularity E.}\label{fig:contours_1}
\end{figure*}

\begin{figure*}[htbp]
\centering
\includegraphics[width=0.95\textwidth]{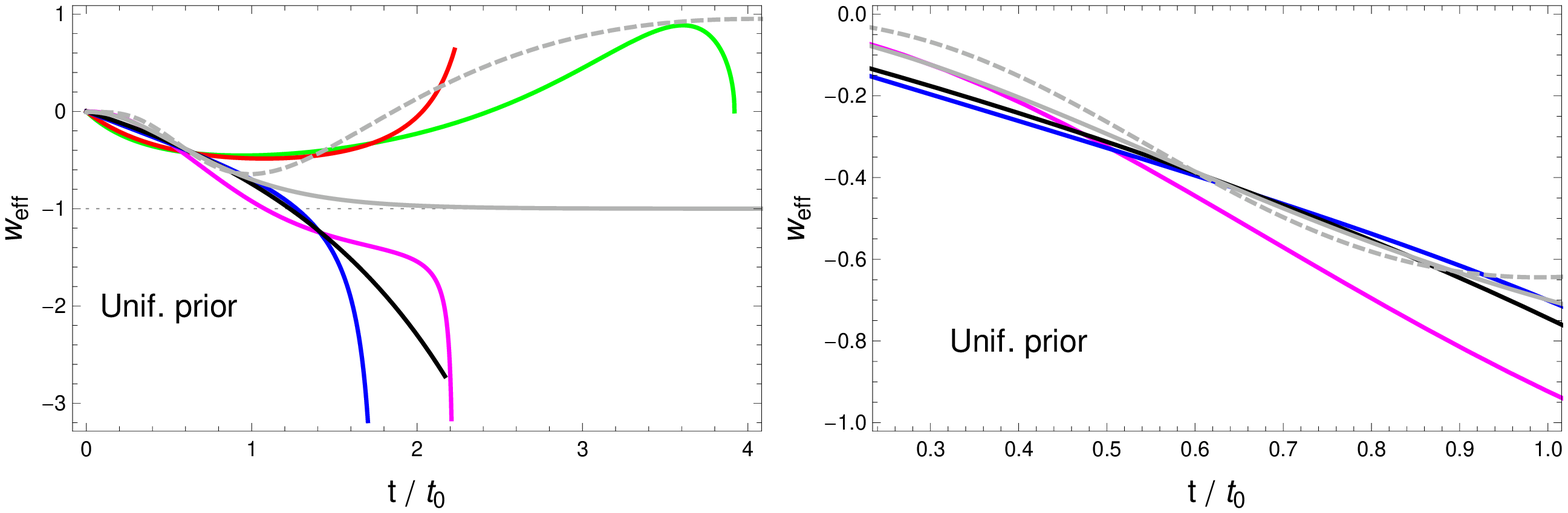}\\
~~~\\
\includegraphics[width=0.95\textwidth]{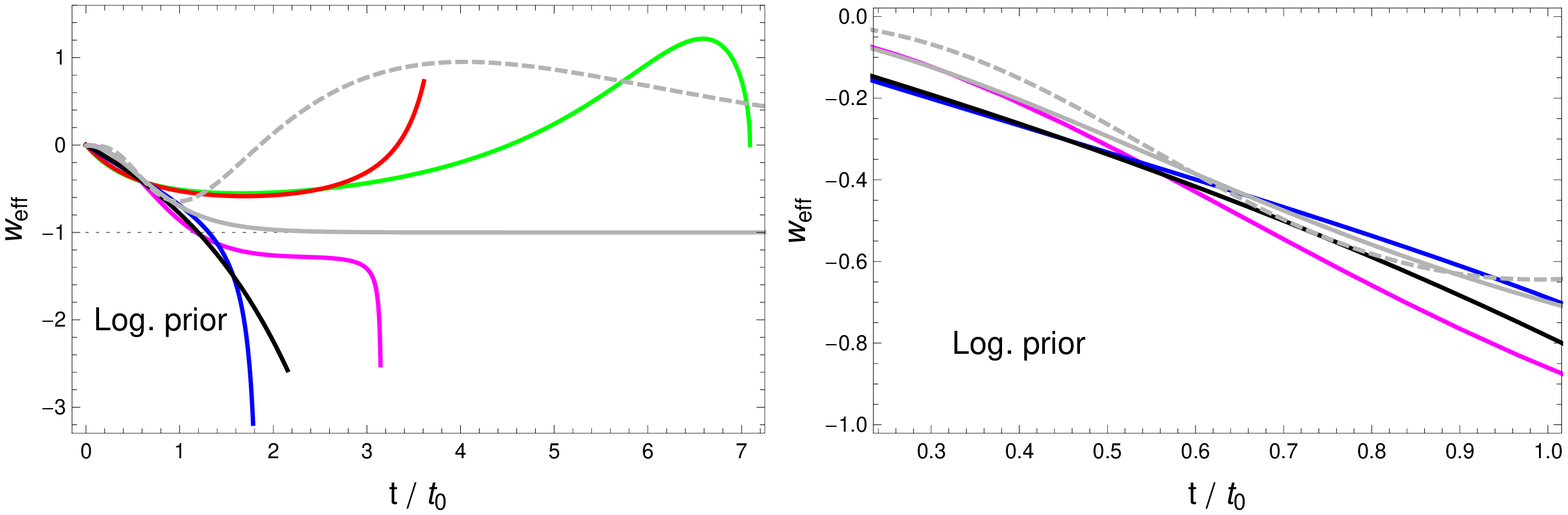}\\
~~~\\
\includegraphics[width=0.95\textwidth]{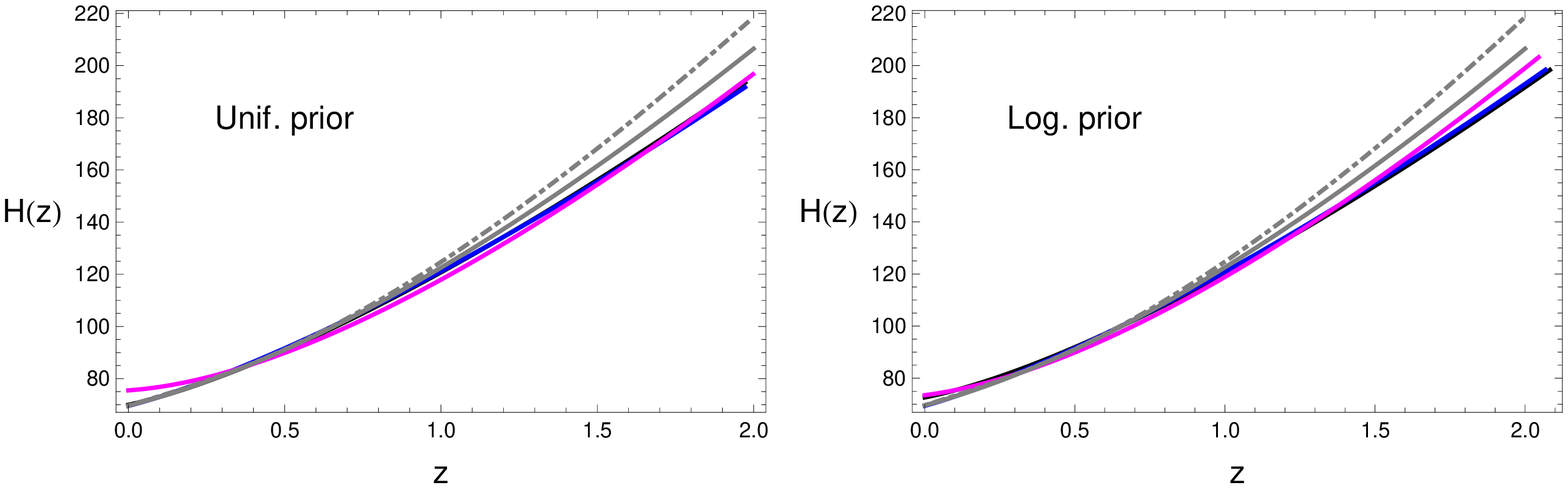}
\caption{\textit{(Top Panel:)} Effective EoS for the combined effect of matter and ``singularity-fluid'' when a uniform prior is applied. \textit{(Center Panel:)} Effective EoS for the combined effect of matter and ``singularity-fluid'' when a logarithmic prior is applied. \textit{(Bottom Panel:)} Rate expansion in terms of redshift, $H(z)$, in the approximate range covered by data. $\Lambda$CDM from Table: solid light grey - CPL from Table: dashed light grey - Singularity A: black - Singularity B: blue - Singularity C: magenta - Singularity D: red - Singularity E: green. (Left: all models for all times - Right: zoom of the best models in the approximate time range covered by data).}
\label{fig:contours_4}
\end{figure*}

\begin{figure*}[htbp]
\centering
\includegraphics[width=0.95\textwidth]{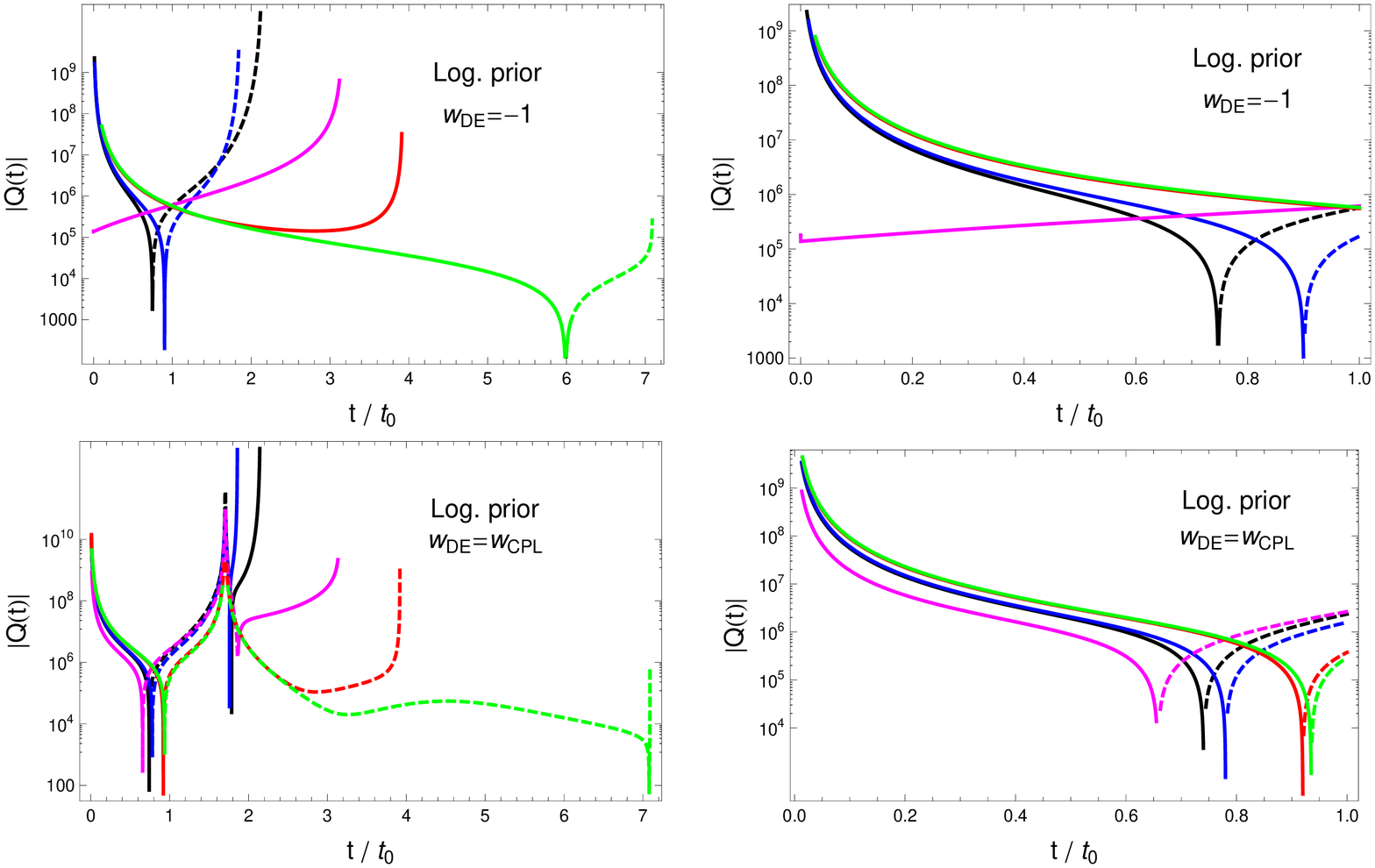}\\
\caption{In this plot we show the interaction $Q$ (normalized to units of $H^3_0/(8\pi G)$ ) derived from our singularity models best fit and assuming a constant dark energy equation of state, $w=-1$. Singularity A: black - Singularity B: blue - Singularity C: magenta - Singularity D: red - Singularity E: green. Left: all models for all times - Right: zoom of the best models in the actual Universe time life range. We used solid (dashed) for positive (negative) $Q$.}\label{fig:interacion}
\end{figure*}

After the datasets introduced in the previous section and the discussed considerations, we have proceeded to run the MCMC chains in order to obtain the confidence regions of each parameterization and, therefore, achieving the main goal of this work, namely, obtaining a lower bound for the time of a future singularity. The results corresponding to our different cases are shown in Fig. \ref{fig:contours_1}, where we display the marginalized contours of the parameters for each model, and in Table \ref{tab:results}. In order to have a further criterium to judge the statistical validity of our models, we have also analyzed, using the same data sets we have described in the previous section, the $\Lambda$CDM model and the Chevallier-Polarski-Linder (CPL) parametrization \citep{CPL}, which is widely used as the most basic generalization of a constant dark energy to a dynamical fluid.

When considering the combination of all the datasets, as can be visually checked from Fig.~\ref{fig:contours_1}, we obtain that the lowest value for the singularity time is achieved for model B and turns out to be $t_{s,{\rm min}}\simeq 1.2$ (at the 2$\sigma$ level), which corresponds to 2.8 Gyrs from today. Remarkably, this lower bound is prior-independent, i.e., it is for both the flat and the logarithmic priors on $\alpha$ that we have used; this helps us to state that such limit is not statistically-biased, but physically compelling. In that regard, it is advantageous to report that the minimum in the $\chi^2$ for models A and B is located, respectively, at $x_{s} \sim 2.18$ and $\sim 1.55$. This does not happen for all other models, whose minima are located  at $\alpha \rightarrow 0$, and are only limited by numerical resolution at $\sim 10^{-4}$. Interestingly, the obtained lower limit for the occurrence of the singularity is shorter than the expected time for the Sun to burn all its fuel (estimated to be 5-7 Gyrs). 

An interesting feature of models A and B is that having the singularity at infinity is excluded at the 1$\sigma$ level, when a uniform prior is assumed; when using a logarithmic prior, model A only lies within the 1$\sigma$ region, while model B still excludes the singularity at infinity at the $1\sigma$ level. We should remember that $t_s=\infty$ corresponds to having a de Sitter universe in the asymptotic future, so for those two models, such a scenario seems to be disfavoured. This highlights that having an asymptotically de Sitter universe in our parameterizations does not necessarily implies being close to a $\Lambda$CDM model. And we also have to point out that when we use a logarithmic prior, which is going to give a better sampling than the uniform prior in the range of very small $\alpha$, such higher bound disappears for model A, but not for model B. Such feature makes this latter model the most interesting among all those we have considered. Remarkably, these two models present a Bayesian evidence which make them equivalent to $\Lambda$CDM from a statistical point of view, i.e., they provide fits as good as those of $\Lambda$CDM, and they are also even better than the widely used CPL parameterization. Notice moreover that the effective equation of state parameter today is close to the one of $\Lambda$CDM. For a more direct comparison of our models with $\Lambda$CDM, in Fig.~\ref{fig:contours_4} we plot the effective equation of state for a total fluid as introduced in Eq. (\ref{eq:weff}) which influences the background dynamics of the universe and the expansion history $H(a)$. However, we need to note that this only happens for the restricted dataset considered in our analysis, while $\Lambda$CDM give a good fit to a much wider variety of cosmological observations, while it is not clear whether the models with singularities will fit all those observations as well as $\Lambda$CDM.

For model C, the possibility of having the singularity at infinity is within the 1$\sigma$ region. The Bayesian evidence in this case is worse than that for models A and B, but still not strongly disfavoured with respect to $\Lambda$CDM. Interestingly the effective equation of state parameter today for this case is substantially lower than for $\Lambda$CDM. Models $D$ and $E$ are strongly disfavoured with respect to the baseline $\Lambda$CDM. Again, these models allow to have $t_s=\infty$ at the 1$\sigma$ level. In these cases, we find that $w_{\mathrm{eff},0}$ is higher than in the $\Lambda$CDM case. Finally, in Fig.~\ref{fig:interacion}, we plot the interaction term given by Eq.~(\ref{Eqbis2}), assuming dark energy equation of state equal to $-1$  and a dynamical one given by the CPL best fit we have found in our analysis, and reported in Table.~\ref{tab:results}.

\section{Conclusion}
\label{Sect6}

In this work we have reconsidered the subject of future cosmological singularities occurring at a finite time. The aim of the work has been to establish a general lower bound for the time of a potential future singularity by using SN Ia, BAO and $H(z)$ data. We have briefly reviewed the cosmological singularities emphasizing the fact that a divergence in a given cosmological parameter does not necessarily implies a singular spacetime. We have then discussed under which conditions a given {\it cosmological} singularity actually corresponds to a {\it singular} spacetime so that we can discern the severity of the different cosmological singularities. Our discussion focused on the geodesic completeness of the spacetime as well as the presence of divergent tidal forces when approaching the singularity.

After this brief theoretical review, we have constructed a set of parameterizations comprising different types of singularities. These parameterizations have been designed so that we recover an early time matter dominated phase that transits to a phase with a future singularity where a given time-derivative of the scale factor diverges, but not the lower ones. We have then run a series of MCMC chains to confront our parameterizations to SN Ia, BAO and $H(z)$ data. The obtained results are then summarized in Table.~\ref{tab:results}. Our main conclusion is that within our family of parameterizations, a potential future singularity cannot be closer to the present time than $\sim 0.2 t_0$, that roughly corresponds to $2.8$ Gyr. We found that the proximity of the singularity to the present time has a mild dependence on the type of singularity for our parameterizations, but we can conclude that in all cases there is a consistent lower bound around $1.2-1.5 t_0$.

Another interesting conclusion that we have found is that, following results from the Bayesian evidence, our parameterizations A and B  provide fits which are not significantly worse than $\Lambda$CDM for the considered datasets. This was not obvious a priori since none of our parameterizations contain $\Lambda$CDM in its parameter space. Hence, as shown in previous references \cite{Lazkoz:2006gp}, a singular scenario can not be discarded right away from tests of the background evolution and the time remaining for the occurrence of a future singularity may be shorter than expected. However, we need to stress that $\Lambda$CDM has become the standard model of cosmology because of its outstanding performance in fitting most cosmological observations, not only the ones considered in our analysis, so that in order to be able to establish a compelling scenario with a future singularity on equal footing as $\Lambda$CDM, we would need to show its ability to fit the rest of cosmological observations, including those sensitive to the perturbations.

\vspace{0.5cm}

{\bf Acknowledgments}: We thank Antonio L. Maroto and  Diego Rubiera-Garcia for fruitful discussions and comments. J.B.J.  acknowledges  the  financial  support  of A*MIDEX project (n. ANR-11-IDEX-0001-02) funded by the ``Investissements d'Avenir'' French Government program, managed by the French National Research Agency (ANR), MINECO (Spain) projects  FIS2011-23000, FIS2014-52837-P, and Consolider-Ingenio MULTIDARK CSD2009-00064. R.L. was supported by the Spanish Ministry of Economy and Competitiveness through research projects FIS2014-57956-P (comprising FEDER funds) and Consolider EPI CSD2010-00064. D.S.G. acknowledges support from a postdoctoral fellowship Ref. SFRH/BPD/95939/2013 by Funda\c{c}\~ao para a Ci\^encia e a Tecnologia (FCT, Portugal) and the support through the research grant UID/FIS/04434/2013 (FCT, Portugal). V.S. is financed by the Polish National Science Center Grant DEC-2012/06/A/ST2/00395.

\end{document}